\def\rN{{\rm NS}}
\def\rR{{\rm R}}
\def\pu(#1){\Phi^{\sigma}(#1)}
\def\pd(#1){\Phi_{\sigma}(#1)}
\def\Hn{{\cal H}^{\rm NS}}
\def\Hr{{\cal H}^{\rm R}}
\def\sn{{\rm sn}\,}
\def\cn{{\rm cn}\,}
\def\dn{{\rm dn}\,}
\def\zn{{\rm zn}\,}
\def\Remark{\medskip\noindent {\sl Remark.}\quad}
\def\H{\cal H}
\def\ze{\zeta}
\def\ve{\varepsilon}
\def\half{\frac{1}{2}}
\def\<{\langle}
\def\>{\rangle}
\def\bu{\bullet}
\def\Z{{\bf Z}}
\def\Zh{\Z+{1\over 2}}
\def\tr{{\rm tr}}
\def\goto#1{\buildrel #1 \over \longrightarrow}
\def\n{\nonumber\\}
\def\id{\mathrm{id}}
\def\a(#1){\alpha_{#1}}
\def\l{\sigma}
\def\Pu(#1,#2){\Phi^{#1}(#2)}
\def\Pd(#1,#2){\Phi_{#1}(#2)}
\def\z(#1){\zeta_{#1}}

\def\hh{{\rm H}}
\def\th{{\rm \Theta}}
\def\h1{{\rm H_1}}
\def\t1{{\rm \Theta_1}}
\def\lr{\leftrightarrow}

\documentclass[12pt]{article}
\usepackage{amssymb,amsfonts,latexsym,epic} 
\begin{document}

\begin{titlepage}

\vspace*{\fill}

\begin{center}
{\large\bf Multi-spin correlation functions for the Z-Invariant Ising model}
\vfill
{\sc J.R.Reyes Mart\'inez}\\[2em]

{\sl Institut f\"ur Theoretische Physik \\
     Freie Universit\"at Berlin \\
     Arnimallee 14, 14195 Berlin \\}
{\rm martinez@physik.fu-berlin.de}

\vfill
{\bf Abstract}

\end{center}

\begin{quote}
Continuing our work \cite{JMR}, where a explicit formula for the
two-point functions of the two dimensional Z-invariant Ising model were found.
I obtain here different results for the higher correlation functions 
and several consistency checks are done.
\end{quote}
\end{titlepage}

\section{Introduction}
\label{sec:Int}

 In this paper I
want to apply the the vertex operator 
method (VOM) for the calculation of different correlations
functions of the  two dimensional Z-invariant Ising
model (IM). Z-invariance is the property that allows to introduce
inhomogeneities in the lattice in conform with the integrability of
the model. In particular when we put all inhomogeneities equal zero we
obtain a homogeneous Ising model in a square lattice. Also by choosing
appropriately the inhomogeneity parameters its is possible to obtain
other regular lattice configurations, like a triangle, checkerboard and
others (see~\cite{AP} for examples). The VOM was introduced by the
Kyoto group in the case of the six-vertex model~\cite{six-vertex}
 and then extended to a
variety of other integrable lattice models. The novelty of this
approach is the treatment from the beginning of the lattice model in
the thermodynamic limit,
these is made possible because of two fundamental concepts: the corner  
transfer matrix (CTM)  of Baxter~\cite{BaxBk} and the lattice 
vertex operator (VO)
introduced by the Kyoto group, see \cite{six-vertex-book} for a
review. Whereas the CTM allows the
calculation of the one-point functions the VOs permits in
connection with the CTM the calculation of higher correlation
functions (CF). Physically they correspond to inserting a
semi-infinite inhomogeneity line at bulk of the lattice. This
inhomogeneity is described by a spectral parameters that enters the
parameterization of the Boltzmann's weights of the model such that they
satisfy some integrability conditions (in the case of the IM they are
called the star-triangle relations), which imply that the diagonal transfer
matrices of the model commute and also allow the possibility of
choosing different spectral parameters in the definition of the usual
transfer matrices~\cite{BaxBk}.

In the paper~\cite{JMR} the author  obtained the two-point functions
in the general Z-invariant IM whose vertex operator algebra was developed
in~\cite{FJMMN}. These were given in forms of
pfaffians of well known elliptic functions over the inhomogeneity
parameters. Because of its importance for the calculation of the 
higher correlation
functions I review this result in Sect.~\ref{sec:spin-spin} (see also
Sect.~\ref{sec:general}) and in
the Appendix I give explicit details of the calculations. After
reviewing some important definitions in Sect.~\ref{sec:fund}, then in
Sect.~\ref{sec:higher} I give a general integral 
formula for all higher
correlations with an even number of spins. Particular
cases of this formula are treated next. Section~\ref{sec:explicit}
contains our principal results, the first part review the general
two-point functions as already mentioned, in Sect.~\ref{sec:even} the
homogeneous CF with adjacent diagonal spins are found in the form of
 generalized
Toeplitz determinants and in Sect.~\ref{sec:diagonal} I find general
homogeneous diagonal four-point functions, obtaining a general determinantal
expression. In Sect.~\ref{sec:odd-point} I consider finally some particular
cases of three-point functions. A conclusion section ends the paper.

\section{Fundamental definitions}
\label{sec:fund}

\subsection{Elliptic parameterization}
\label{sec:BW}

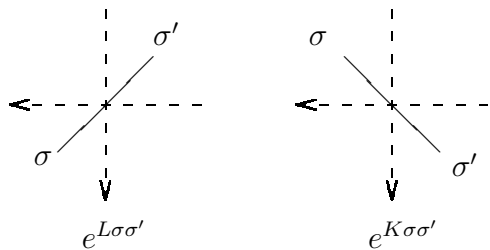
\begin{figure}[ht]
\begin{center}
  \setlength{\unitlength}{0.0125in}
    \begin{picture}(209,115)(0,-10)
\drawline(20,40)(60,80)
\dashline{4.000}(40,100)(40,20)
\drawline(38.000,28.000)(40.000,20.000)(42.000,28.000)
\dashline{4.000}(80,60)(0,60)
\drawline(8.000,62.000)(0.000,60.000)(8.000,58.000)
\dashline{4.000}(200,60)(120,60)
\drawline(128.000,62.000)(120.000,60.000)(128.000,58.000)
\dashline{4.000}(160,100)(160,20)
\drawline(158.000,28.000)(160.000,20.000)(162.000,28.000)
\drawline(180,40)(140,80)
\put(150,0){\makebox(0,0)[lb]{\raisebox{0pt}[0pt][0pt]{\shortstack[l]{{\rm
$e^{K\l\l'}$}}}}}
\put(30,0){\makebox(0,0)[lb]{\raisebox{0pt}[0pt][0pt]{\shortstack[l]{{\rm
$e^{L\l\l'}$}}}}}
\put(185,30){\makebox(0,0)[lb]{\raisebox{0pt}[0pt][0pt]{\shortstack[l]{{\rm
$\l'$}}}}}
\put(60,85){\makebox(0,0)[lb]{\raisebox{0pt}[0pt][0pt]{\shortstack[l]{{\rm
$\l'$}}}}}
\put(125,85){\makebox(0,0)[lb]{\raisebox{0pt}[0pt][0pt]{\shortstack[l]{{\rm
$\l$}}}}}
\put(10,35){\makebox(0,0)[lb]{\raisebox{0pt}[0pt][0pt]{\shortstack[l]{{\rm
$\l$}}}}}
     \end{picture}
       \caption{Boltzmann Weights for the IM}
        \label{fig:BW}
   \end{center}
\end{figure}

Let us begin by introducing a number of definitions for the 
Z-invariant IM; the model is defined initially on a square lattice
rotated by 45 degrees, the spin variables taking values $+1$ or $-1$
live on the vertices of the lattice the coupling being between nearest
neighbors. Assume first two different interaction energies (measured 
in $\mathit{k}T$ units) with $K >0$ in
the NW-SO direction and $L > 0$ in the NO-SW direction between two spins
as in the figure~\ref{fig:BW}

In the figure the two lines indicate one horizontal $u_H$ and
one vertical $u_V$ spectral parameters, their difference $u = u_H -
u_V$ is used to code the couplings energies under the critical
temperature by mean of Jacobian elliptic functions,
\begin{eqnarray}
\label{eq:bw}
&&\sinh(2K)= -i\sn iu ,\quad \cosh(2K)=\cn iu,\n
&&\sinh(2L)=  i k^{-1}{\rm ns}\, iu,\quad \cosh(2L)=ik^{-1} {\rm ds}\,iu,
\end{eqnarray}
where $\sn u$ is a Jacobian  elliptic
function with half-periods  $I$, $iI'$, $k$ is the
modulus, $0< u< I'$ (see e.g. \cite{BaxBk}),
and Glashier`s notation for the Jacobian elliptic functions is
used, $\mathrm{ns}\,u :=1/\sn u$, 
$\mathrm{ds}\, u := \dn u/\sn u$. These parameterization already
introduced by Onsager is a consequence of the integrability
conditions, in the case of the IM the so called star-triangle
relations, which imply  as well known, that  
diagonal matrices  commute when the difference of the spectral
parameters from site to site remain constant (see pag.~374 in 
Baxter's book). The module $k = (\sinh(2K)\sinh(2L))^{-1}$ does not
depend on the spectral parameter and tends to $1$ when the systems
goes critical. It is possible to write a similar parameterization
for temperatures higher than the critical but it will not be necessary
due to the duality between order and disorder operators to be
introduced below, which imply that the disorder CF  under
the critical temperature correspond to the order CF above it.

\subsection{Vertex Operator Algebra}
\label{sec:VOA}

\begin{figure}[htbp]
  \setlength{\unitlength}{0.0125in}
\begin{picture}(200,195)(0,-10)
\drawline(100,0)(40,60)
\drawline(180,80)(120,140)
\drawline(180,0)(80,100)
\drawline(180,80)(100,0)
\drawline(180,160)(20,0)
\dashline{4.000}(180,140)(120,140)
\drawline(128.000,142.000)(120.000,140.000)(128.000,138.000)
\dashline{4.000}(180,100)(80,100)
\drawline(88.000,102.000)(80.000,100.000)(88.000,98.000)
\dashline{4.000}(180,60)(40,60)
\drawline(48.000,62.000)(40.000,60.000)(48.000,58.000)
\dashline{4.000}(180,20)(0,20)
\drawline(8.000,22.000)(0.000,20.000)(8.000,18.000)
\dashline{4.000}(40,60)(40,0)
\drawline(38.000,8.000)(40.000,0.000)(42.000,8.000)
\dashline{4.000}(80,100)(80,0)
\drawline(78.000,8.000)(80.000,0.000)(82.000,8.000)
\dashline{4.000}(120,140)(120,0)
\drawline(118.000,8.000)(120.000,0.000)(122.000,8.000)
\dashline{4.000}(160,180)(160,0)
\drawline(158.000,8.000)(160.000,0.000)(162.000,8.000)
\put(25,65){\makebox(0,0)[lb]{\raisebox{0pt}[0pt][0pt]{\shortstack[l]{{\rm
$+$}}}}}
\put(65,105){\makebox(0,0)[lb]{\raisebox{0pt}[0pt][0pt]{\shortstack[l]{{\rm
$+$}}}}}
\put(105,140){\makebox(0,0)[lb]{\raisebox{0pt}[0pt][0pt]{\shortstack[l]{{\rm
$+$}}}}}
\put(20,170){\makebox(0,0)[lb]{\raisebox{0pt}[0pt][0pt]{\shortstack[l]{{\rm
$A_\rN$}}}}}
\end{picture}
\begin{picture}(200,195)(0,-10)
\drawline(280,40)(240,0)
\drawline(280,120)(160,0)
\drawline(160,0)(120,40)
\drawline(240,0)(160,80)
\drawline(280,120)(240,160)
\drawline(280,40)(200,120)
\dashline{4.000}(280,140)(220,140)
\drawline(228.000,142.000)(220.000,140.000)(228.000,138.000)
\dashline{4.000}(280,100)(180,100)
\drawline(188.000,102.000)(180.000,100.000)(188.000,98.000)
\dashline{4.000}(280,60)(140,60)
\drawline(148.000,62.000)(140.000,60.000)(148.000,58.000)
\dashline{4.000}(280,20)(100,20)
\drawline(108.000,22.000)(100.000,20.000)(108.000,18.000)
\dashline{4.000}(140,60)(140,0)
\drawline(138.000,8.000)(140.000,0.000)(142.000,8.000)
\dashline{4.000}(180,100)(180,0)
\drawline(178.000,8.000)(180.000,0.000)(182.000,8.000)
\dashline{4.000}(220,140)(220,0)
\drawline(218.000,8.000)(220.000,0.000)(222.000,8.000)
\dashline{4.000}(260,180)(260,0)
\drawline(258.000,8.000)(260.000,0.000)(262.000,8.000)
\put(110,45){\makebox(0,0)[lb]{\raisebox{0pt}[0pt][0pt]{\shortstack[l]{{\rm
$+$}}}}}
\put(150,85){\makebox(0,0)[lb]{\raisebox{0pt}[0pt][0pt]{\shortstack[l]{{\rm
$+$}}}}}
\put(190,125){\makebox(0,0)[lb]{\raisebox{0pt}[0pt][0pt]{\shortstack[l]{{\rm
$+$}}}}}
\put(230,165){\makebox(0,0)[lb]{\raisebox{0pt}[0pt][0pt]{\shortstack[l]{{\rm
$+$}}}}}
\put(130,170){\makebox(0,0)[lb]{\raisebox{0pt}[0pt][0pt]{\shortstack[l]{{\rm
$A_\rR$}}}}}
\end{picture}
 \caption{The CTMs for the IM}
\label{fig:CTM}
\end{figure}
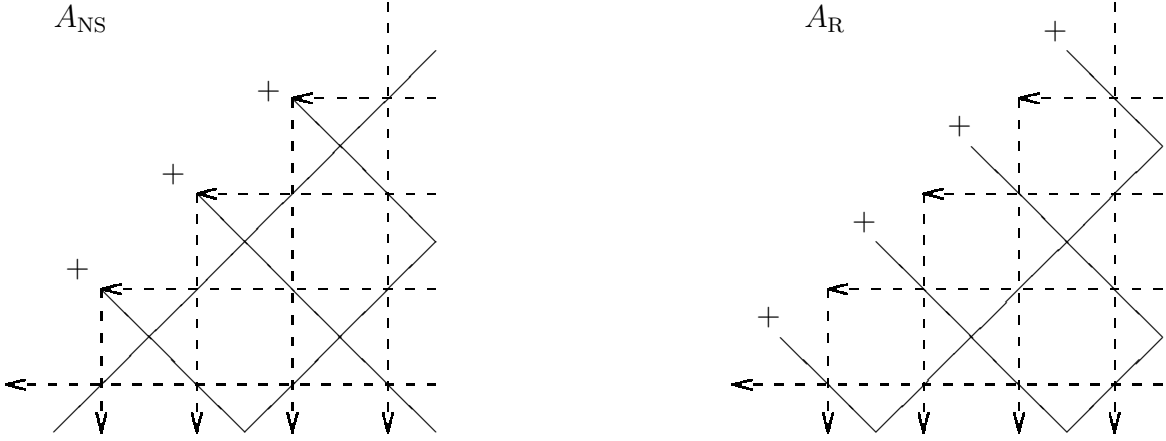

I review now the fundamental results of the paper~\cite{FJMMN}, 
where the vertex operator algebra for the IM was found.
Dividing
the diagonal lattice in four equal quadrants to apply
Baxter's CTM method we have two cases according
to wether the central point belongs or not to the lattice;
this defines {\em two} CTM's~Fig.\ref{fig:CTM} that act on the
two sectors of the model, $\rN$ and $\rR$,
respectively (in analogy to conformal field theory).
Diagonalising the adjoint action of the CTM Hamiltonians 
we can describe the two sectors as Fock spaces (called
$\Hn$ and $\Hr$)  of
two irreducible representations of free fermions (apart from a
different normalization).
We have:
\begin{eqnarray*}
&&[D^\rN,\phi^\rN_r]=-2r\phi^\rN_r,\quad
[D^\rR,\phi^\rR_r]=-2r\phi^\rR_r ,
\end{eqnarray*}
where 
$r\in\Zh$ for the $NS$ sector,
$r\in\Z$ for the $R$ sector and
$D^\rN$, $D^\rR$ are the respective CTM hamiltonians.

There are two vertex operators (VOs in the following)
depending on which sector they intertwine:
\begin{eqnarray*}
&&\Phi_{\rN}^{\rR}(\zeta):\Hn\goto{} \Hr, \\
&&\Phi_{\rR}^{\rN}(\zeta):\Hr\goto{} \Hn,
\end{eqnarray*}

The Vertex Operators (VO) satisfy the following elegant intertwining
relations with the fermions
\begin{eqnarray}
\label{eq:inter1}
\Phi^\sigma(x\ze)\phi^\rR(z)&=&
\phi^\rN(z)f(z\ze^2)\Phi^{-\sigma}(x\ze), \n
\Phi_\sigma(x\ze)\phi^\rN(z)&=&
f(z\ze^2)\phi^\rR(z)\Phi_{-\sigma}(x\ze), 
\end{eqnarray}
\begin{eqnarray}
\label{eq:inter1.2}
\l\Phi^\l(x\ze)&=&i\psi_1^\rN(\ze)\Phi^{-\l}(x\ze), \n
\l\Phi_\l(x\ze)&=&\psi_1^\rR(\ze)\Phi_{\l}(x\ze).
\end{eqnarray}
Here we use
the simplified notation of \cite{FJMMN} 
$\Phi^{\rN,\sigma}_\rR(\zeta)=\pu(\ze)$,
$\sigma$ being equal to $+1$ or $-1$ distinguishes the
two sub-sectors (with even or odd parity)
of the $\rN$ sector and
$\Phi_{\rN,\sigma}^{\rR}(\zeta)=\pd(\ze)$, also we define
$[z] := f(z)= -\sqrt{k}{\rm sn} v$ with $z=\exp(i\pi v/I)$
and we have introduced the generating functions
\begin{eqnarray}
\label{eq:phi}
&&\phi^\rN(z)=\sum_{r\in \Zh}\phi^\rN_r z^{-r},\quad 
\phi^\rR(z)=\sum_{s\in \Z}\phi^\rR_s z^{-s},
\end{eqnarray}
and
\begin{eqnarray}
\label{eq:psi-phi}
\psi_1^\rN(\zeta) =
\oint{dz\over 2\pi i z} [z]_{\rN}\phi^\rN(z/\zeta^2),\qquad
\psi_1^\rR(\zeta) =
\oint{dz\over 2\pi i z} [z]_{\rR}\phi^\rR(z/\zeta^2),
\end{eqnarray}
where
\begin{eqnarray*}
&&[z]_{\rN}=\sqrt{2I k\over \pi}{\rm cn} v
=\sqrt{2\pi \over kI}\sum_{r\in \Zh}\eta^{-1}_r z^r,\n
&&[z]_{\rR}=\sqrt{ 2I\over \pi}{\rm dn} v
=\sqrt{2\pi \over I}\sum_{s\in \Z}\eta^{-1}_s z^s, \n
&&\eta_r =x^{2r} + x^{-2r},\qquad x =\exp(-\frac{\pi I'}{2I}).
\end{eqnarray*}
By crossing symmetry i.e. $u\rightarrow u - I'$ 
or in the multiplicative language $x\ze\rightarrow \ze$
we can obtain  (using also the transformation
$f(x^2z)=1/f(z)$) similar relations with the parameter
of the VO replaced with $\ze$ but in the calculation of the CF
in the next sections I will need only the expressions above.
Finally the VOs satisfy the following unitary relations:
\begin{eqnarray}
&&\sum_\sigma
\Phi_{\sigma}(x\zeta)
\Phi^\sigma(\zeta)=g^{\rm R}\times {\rm id}_{\H^\rR},
\label{eq:inverse1}\\
&&
\Phi^{\sigma}(x\zeta)\Phi_{\sigma}(\zeta)
=g^{\rm NS}\times {\rm id}_{\H^{\rN,\sigma}},
\label{eq:inverse2}
\end{eqnarray}
where the scalars $g^{\rm R}$ and $g^{\rm NS}$ can be given
as infinite products in the variable $x$ but as we will not
need them for the calculation of the CF we leave them undefined,
similarly the expectation values of the VOs in the respective fermion
basis are known but will not be needed 
(see~\cite{FJMMN} for their explicit values).

\section{Even correlation functions}
\label{sec:spin-spin}

In this section I will give a general multi-integral
expression for the CF with an
even number of spins. The case two spin operators was considered
in~\cite{JMR} where the integration was carried explicitly. This
result will be remembered in the  Sect.~\ref{sec:general}. We start with
an easy example an proceed in order of difficulty.

\subsection{Nearest diagonal neighbors CF}
\label{sec:nearest}

\begin{figure}[htbp]
  \begin{center}
\setlength{\unitlength}{0.0125in}
\begin{picture}(280,120)
\dashline{4.000}(0,40)(120,40)
\drawline(8.000,38.000)(0.000,40.000)(8.000,42.000)
\dashline{4.000}(160,37)(280,37)
\drawline(168.000,35.000)(160.000,37.000)(168.000,39.000)
\dashline{4.000}(0,80)(120,80)
\drawline(8.000,78.000)(0.000,80.000)(8.000,82.000)
\dashline{4.000}(160,83)(280,83)
\drawline(168.000,81.000)(160.000,83.000)(168.000,85.000)
\dashline{4.000}(40,0)(40,120)
\drawline(38.000,8.000)(40.000,0.000)(42.000,8.000)
\dashline{4.000}(80,0)(80,120)
\drawline(78.000,8.000)(80.000,0.000)(82.000,8.000)
\dashline{4.000}(200,0)(200,120)
\drawline(198.000,8.000)(200.000,0.000)(202.000,8.000)
\dashline{4.000}(240,0)(240,120)
\drawline(238.000,8.000)(240.000,0.000)(242.000,8.000)
\drawline(100,60)(60,20)(20,60)
\drawline(100,60)(60,100)(20,60)
\drawline(180,20)(220,55)(260,20)
\drawline(180,100)(220,65)(260,100)

\put(40,-15){\makebox(0,0)[lb]{\raisebox{0pt}[0pt][0pt]{\shortstack[l]{{\rm
$\ze_1$}}}}}
\put(80,-15){\makebox(0,0)[lb]{\raisebox{0pt}[0pt][0pt]{\shortstack[l]{{\rm
$\ze_2$}}}}}
\put(200,-15){\makebox(0,0)[lb]{\raisebox{0pt}[0pt][0pt]{\shortstack[l]{{\rm
$\ze_1$}}}}}
\put(240,-15){\makebox(0,0)[lb]{\raisebox{0pt}[0pt][0pt]{\shortstack[l]{{\rm
$\ze_2$}}}}}

\put(10,60){\makebox(0,0)[lb]{\raisebox{0pt}[0pt][0pt]{\shortstack[l]{{\rm
$\l$}}}}}
\put(105,60){\makebox(0,0)[lb]{\raisebox{0pt}[0pt][0pt]{\shortstack[l]{{\rm
$\l'$}}}}}
\put(210,50){\makebox(0,0)[lb]{\raisebox{0pt}[0pt][0pt]{\shortstack[l]{{\rm
$\l$}}}}}
\put(210,60){\makebox(0,0)[lb]{\raisebox{0pt}[0pt][0pt]{\shortstack[l]{{\rm
$-\l$}}}}}

\end{picture}
\end{center}
\caption{Order-Order and Disorder-Disorder CF for nearest diagonal neighbors}
  \label{fig:spin-spin}
\end{figure}
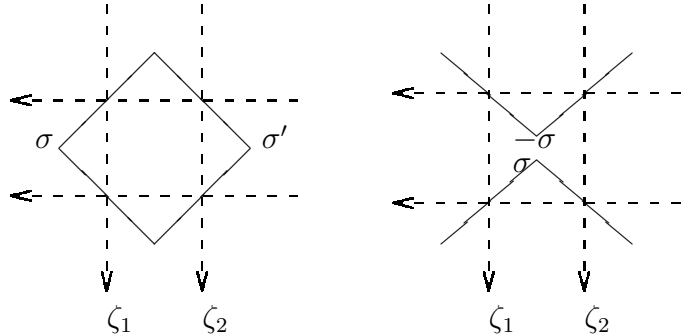

Now I review the calculation of the two-point functions. For
simplicity let us consider first the case for  nearest diagonal 
order-order  and disorder-disorder CF  Fig.~\ref{fig:spin-spin}.
In term of the VO they are respectively
\begin{eqnarray}
&&\frac{\sum_{\l\l'}\l\l' \tr_{\H^\rN}
\left(x^{2D^\rN}\Phi^{\l'}(x\ze_2)\Phi_\l(x\ze_1)
\Phi^\l(\ze_1)\Phi_{\l'}(\ze_2)\right)}
{g^\rN g^\rR\tr_{\H^\rN}\bigl(x^{2D^\rN}\bigr)},\label{eq:order}\\
&&\frac{\sum_\l\tr_{\H^\rR}
\left(x^{2D^\rR}\Phi_{\l}(x\ze_2)\Phi^\l(x\ze_1)
\Phi_{-\l}(\ze_1)\Phi^{-\l}(\ze_2)\right)} 
{g^\rN g^\rR\tr_{\H^\rR}\bigl(x^{2D^\rR}\bigr)}.
\label{eq:disorder}
\end{eqnarray}
Consider as an example~(\ref{eq:disorder}); the reader can look
at App.~B in \cite{FJMMN} for a explicit determination of 
(\ref{eq:order}). We transform the four VO under the trace symbol
in terms of fermions in the following way
\begin{eqnarray}
\label{eq:4d}
&&\sum_{\l}\Phi_{\l}(x\ze_{2})\Phi^{\l}(x\ze_{1})
\Phi_{-\l}(\ze_1)\Phi^{-\l}(\ze_{2})= \nonumber \\
&&\sum_{\l}\l\Phi_{\l}(x\ze_{2})\l\Phi^{\l}(x\ze_{1})
\Phi_{-\l}(\ze_1)\Phi^{-\l}(\ze_{2})= \nonumber \\
&&\sum_{\l}\psi_1^{\rm R}(\ze_l)\Phi_{\l}(x\ze_{2}) i 
\psi_1^{\rm NS}(\ze_1)\Phi^{-\l}(x\ze_{1})
\Phi_{-\l}(\ze_1)\Phi^{-\l}(\ze_{2})= \nonumber \\ 
&&\psi_1^{\rm R}(\ze_2)\oint_{w_1}\, [w_1 1]_\rN
\phi^\rR(w_1)\, [w_1 2]\times\sum_{-\l} 
\Phi_{-\l}(x\ze_{2})\Phi^{-\l}(x\ze_{1})
\Phi_{-\l}(\ze_1)\Phi^{-\l}(\ze_{2})= \nonumber \\
&&ig^{\rm NS}g^{\rm R}\oint_{w_2}
\, [w_{2}2]_\rR\phi^\rR(w_{2})\times 
\oint_{w_1}\, [w_11]_\rN\, [w_12]
\phi^\rR(w_1)=  \nonumber \\
&&ig^{\rm NS}g^{\rm R}\oint_{w_{2}}\oint_{w_1}
[w_{1} 2]_{\rm R}[w_1 1]_{\rm NS}[w_1 2]
\phi^\rR(w_{2})\phi^\rR(w_1),
\end{eqnarray}
where we changed the variables to $w_i = z_i/\ze^2_i$ and used
the notation $\oint_i =\oint\frac{dw_i}{2\pi i w_i}$,
$[w_ij]=[w_i\ze^2_j]$.
Following substitutions were then made,
in the second line  the identity $\l^2 = 1$, in
the third line eqs.~(\ref{eq:inter1.2}), 
the intertwining relation (\ref{eq:inter1}) and
the second eq. in (\ref{eq:psi-phi}) in the fourth line and
finally the unitarity conditions for the VOs.

It only remain to take the trace of the fermion operators, which we
state as a simple lemma, 
\begin{equation}
\label{eq:lemma}
\frac{\tr_{\Hr}\left(x^{2D^\rR}\phi^\rR(w_2)\phi^\rR(w_1)\right)}
{\tr_{\Hr}\bigl(x^{2D^\rR}\bigr)}=
\delta^\rR(x^2w_2/w_1).
\end{equation}
With the aid of this formula and the relation 
$[x^{-2}z]_\rR[z] =-i[z]_\rN$
we are leave with an integral expression
for the disorder-disorder CF,
\begin{equation}
\label{eq:dis-dis}
\oint_w[w 1]_\rN[w 2]_\rN.
\end{equation}
Remembering that $w =\exp(i\pi v/I)$ and $\ze =\exp(-\pi\alpha/2I)$
the last expression can be written as:
\[
\frac{k}{\pi}\int_{-I}^{I}\!\!dv\;\cn(v - i\a(1))\cn(v - i\a(2))=
-\frac{2I}{\pi}\left(\frac{k'}{k}\right)^{\frac{1}{2}}
\frac{\t1'(i\a(12)))}{\hh(i\a(12))},
\]
where $\hh$, $\th$, $\h1$ and $\t1$ (see e.g.\cite{grad}) are the
Jacobi theta functions and $\a(12):=\a(1)-\a(2)$. The prime denotes
the derivative under the argument. This integration formula appears in
other disguise in the Appendix, see eq.~(\ref{app:cn}).

\subsection{Two-point functions with an arbitrary number of
  inhomogeneities}
\label{sec:arbitrary}

Let us call the general order-order and disorder-disorder CF by
$g^{\sigma}_{2n}(\a(1),\a(2),\ldots,\a(2n))$ and
$g^{\mu}_{2n}(\a(1),\a(2),\ldots,\a(2n))$ respectively (when we
don't want to emphasize the spectral parameters $g^\l_{2n}$ and
$g^\mu_{2n}$ simply),
where $\ze_j=\exp(-\pi\a(j)/2I)$,  then we have as simple
generalizations of the trace formulas from Sect.~\ref{sec:nearest}:
\begin{eqnarray}
  \label{eq:order1}
g^{\sigma}_{2n}(\a(1),\a(2),\ldots,\a(2n))=
\frac{\sum_{\l;\mu_i;\l'}\l\l'\tr_{\rN}\left(x^{2D^{\rN}}
F(\l,\l';\mu_i;\z(1),\ldots,\z(2n))\right)}
{(g^{\rN}g^{\rR})^n\tr_{\rN}\left( x^{2D^{\rN}}\right)},
\end{eqnarray}
\begin{eqnarray*}
&&F(\l,\l';\mu_i;\z(1),\ldots,\z(2n))= \n
&&\Pu(\l',x\z(2n))\Pd(\mu_{n-1},x\z(2n-1))
\cdots\Pu(\mu_{1},x\z(2))\Pd(\l,x\z(1))\Pu(\l,\z(1))\Pd(\mu_{1},\z(2))
\cdots \Pu(\mu_{n-1},\z(2n-1))\Pd(\l',\z(2n)),
\end{eqnarray*}
where the subscript $\mu_i$ in the sum above means the sum for all
$1\leq i\leq n-1$. Similarly for the disorder-disorder case:
\begin{eqnarray}
  \label{eq:disorder1}
g^{\mu}_{2n}(\a(1),\a(2),\ldots,\a(2n))=
\frac{\sum_{\mu_i}\tr_{\rR}\left(x^{2D^{\rR}}
G(\mu_i;\z(1),\ldots,\z(2n))\right)}
{(g^{\rN}g^{\rR})^n\tr_{\rR}\left( x^{2D^{\rR}}\right)},
\end{eqnarray}
\begin{eqnarray*}
&&G(\mu_i;\z(1),\ldots,\z(2n))= \n
&&\Pd(\mu_{n},x\z(2n))\Pu(\mu_{n},x\z(2n-1))
\cdots\Pd(\mu_1,x\z(2))\Pu(\mu_1,x\z(1))
\Pd(-\mu_1,\z(1))\Pu(-\mu_{1},\z(2))
\cdots \Pd(-\mu_{n},\z(2n-1))\Pu(-\mu_n,\z(2n)),
\end{eqnarray*}
with $1\leq i\leq n$. Consider the disorder CF
(the other case is similar, see~\cite{JMR} for the differences). We
see that if we begin in the middle of the operator expression
$G(\mu_i;\z(1),\ldots,\z(2n))$ we have the same product of 
four vertex operators (\ref{eq:4d}) 
appearing  before, thus we can substitute them by
a pair of fermion operators and move them to the left by means of the
intertwining property~(\ref{eq:inter1}). The result is, apart from
a product of elliptic functions depending on the spectral parameters,
of the same form but with four vertex operators less as
before. Trading in this way each time four vertex operators by two
fermions operators and using the following antisymmetry property
of the fermion algebra under the integrations, i.e.,
\begin{eqnarray*}
&&\oint_{w_{l}}\oint_{w_i}
[w_{l} l]_{\rm R}[w_i i]_{\rm NS}[w_i l]
[\phi^\rR(w_{l}),\phi^\rR(w_i)]_+ = \\
&&\oint_{w_{l}}\oint_{w_i}
[w_{l} l]_{\rm R}[w_i i]_{\rm NS}[w_i l]
(\delta^\rR(x^2w_l/w_i)+\delta^\rR(x^{-2}w_l/w_i))=0,
\end{eqnarray*}
one arrives at the expression (with the fermions ordered from left to
right according to increasing indexes)
\begin{eqnarray}
\label{eq:product}
 (-ig^{\rm NS}g^{\rm R})^{n}
\left\{\prod_{1\leq i \leq 2n}\oint_{w_i}\right\}
[w_1]_{\rN}[w_2]_{\rR}\cdots
[w_{2n-1}]_{\rN}[w_{2n}]_{\rR}\times\!\! 
\prod_{1\leq i < j \leq 2n}[w_ij]\prod_{i=1}^{2n}\phi^\rR(w_i).
\end{eqnarray}
The corresponding formula for the order-order CF is obtained
by simply interchanging the $\rN$ with $\rR$ representation and
vice-versa. We continue with the analysis of this expression in the
Section~\ref{sec:explicit}.

\subsection{Higher even correlation functions}
\label{sec:higher}

By the same method of calculation of the two-point functions explained
above the generalization to arbitrary higher even CF is
straightforward. We consider the CF with only order operators the
purely disorder case can be obtained in the same way as before
interchanging representations.
Let us call the general 2n-CF $G_{2n}$. Define first some parameters:
\begin{itemize}
\item $2n$ spins $\l_i$ for $i$ between $1$ and $2n$ at arbitrary
  positions in a diagonal ordered from left to right;
\item $2N$ spectral parameters $\ze_i$ for $i$ between 1 and $2N$,
$N$ being a positive integer;
\item  $2n$ integers $a_i \in {1,2,\ldots, 2N}$ with 
$a_1=1<a_2<a_3<\cdots<a_{2n}=2N$, such that for odd $i$ the spectral
parameter $\ze_i$ is adjacent to the right of the spin $\l_i$ and
when $i$ is even  $\ze_i$ is adjacent to the left  of the spin $\l_i$;
\item A set ${\cal W}$ of
$ L :=\sum_{i=1}^{n}(a_{2i}-a_{2i-1}+1)$ complex parameters $w_j$,
note that $L$ is an even integer, ordered in the following way: 

${w_{a_1}, w_{a_1+1}, \ldots, w_{a_2};\quad 
w_{a_3}, w_{a_3+1}, \ldots, w_{a_4};\quad  \ldots\quad; 
w_{a_{2n-1}}, w_{a_{2n-1}+1}, \ldots, w_{a_{2n}}}$.
\end{itemize}
Consider the following example for a 6-point CF:
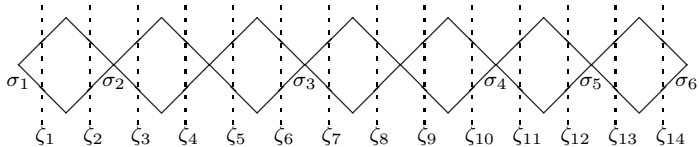
\begin{figure}[ht]
\begin{center}
 \setlength{\unitlength}{0.0125in}
  \begin{picture}(300,60)
\drawline(0,30)(20,50)(40,30)(60,50)(80,30)(100,50)(120,30)(140,50)
(160,30)(180,50)(200,30)(220,50)(240,30)(260,50)(280,30)
(260,10)(240,30)(220,10)(200,30)(180,10)(160,30)(140,10)(120,30)
(100,10)(80,30)(60,10)(40,30)(20,10)(0,30)
\dashline{2}(10,5)(10,55)
\dashline{2}(30,5)(30,55)
\dashline{2}(50,5)(50,55)
\dashline{2}(70,5)(70,55)
\dashline{2}(90,5)(90,55)
\dashline{2}(110,5)(110,55)
\dashline{2}(130,5)(130,55)
\dashline{2}(150,5)(150,55)
\dashline{2}(170,5)(170,55)
\dashline{2}(190,5)(190,55)
\dashline{2}(210,5)(210,55)
\dashline{2}(230,5)(230,55)
\dashline{2}(250,5)(250,55)
\dashline{2}(270,5)(270,55)
\put(7,0){\makebox(0,0)[lb]{\raisebox{0pt}[0pt][0pt]{\shortstack[l]{{\rm
${_{\ze_1}}$}}}}}
\put(27,0){\makebox(0,0)[lb]{\raisebox{0pt}[0pt][0pt]{\shortstack[l]{{\rm
${_{\ze_2}}$}}}}}
\put(47,0){\makebox(0,0)[lb]{\raisebox{0pt}[0pt][0pt]{\shortstack[l]{{\rm
${_{\ze_3}}$}}}}}
\put(67,0){\makebox(0,0)[lb]{\raisebox{0pt}[0pt][0pt]{\shortstack[l]{{\rm
${_{\ze_4}}$}}}}}
\put(87,0){\makebox(0,0)[lb]{\raisebox{0pt}[0pt][0pt]{\shortstack[l]{{\rm
${_{\ze_5}}$}}}}}
\put(107,0){\makebox(0,0)[lb]{\raisebox{0pt}[0pt][0pt]{\shortstack[l]{{\rm
${_{\ze_6}}$}}}}}
\put(127,0){\makebox(0,0)[lb]{\raisebox{0pt}[0pt][0pt]{\shortstack[l]{{\rm
${_{\ze_7}}$}}}}}
\put(147,0){\makebox(0,0)[lb]{\raisebox{0pt}[0pt][0pt]{\shortstack[l]{{\rm
${_{\ze_8}}$}}}}}
\put(167,0){\makebox(0,0)[lb]{\raisebox{0pt}[0pt][0pt]{\shortstack[l]{{\rm
${_{\ze_9}}$}}}}}
\put(187,0){\makebox(0,0)[lb]{\raisebox{0pt}[0pt][0pt]{\shortstack[l]{{\rm
${_{\ze_{10}}}$}}}}}
\put(207,0){\makebox(0,0)[lb]{\raisebox{0pt}[0pt][0pt]{\shortstack[l]{{\rm
${_{\ze_{11}}}$}}}}}
\put(227,0){\makebox(0,0)[lb]{\raisebox{0pt}[0pt][0pt]{\shortstack[l]{{\rm
${_{\ze_{12}}}$}}}}}
\put(247,0){\makebox(0,0)[lb]{\raisebox{0pt}[0pt][0pt]{\shortstack[l]{{\rm
${_{\ze_{13}}}$}}}}}
\put(267,0){\makebox(0,0)[lb]{\raisebox{0pt}[0pt][0pt]{\shortstack[l]{{\rm
${_{\ze_{14}}}$}}}}}
\put(-5,23){\makebox(0,0)[lb]{\raisebox{0pt}[0pt][0pt]{\shortstack[l]{{\rm
${_{\l_1}}$}}}}}
\put(35,23){\makebox(0,0)[lb]{\raisebox{0pt}[0pt][0pt]{\shortstack[l]{{\rm
${_{\l_2}}$}}}}}
\put(115,23){\makebox(0,0)[lb]{\raisebox{0pt}[0pt][0pt]{\shortstack[l]{{\rm
${_{\l_3}}$}}}}}
\put(195,23){\makebox(0,0)[lb]{\raisebox{0pt}[0pt][0pt]{\shortstack[l]{{\rm
${_{\l_4}}$}}}}}
\put(235,23){\makebox(0,0)[lb]{\raisebox{0pt}[0pt][0pt]{\shortstack[l]{{\rm
${_{\l_5}}$}}}}}
\put(275,23){\makebox(0,0)[lb]{\raisebox{0pt}[0pt][0pt]{\shortstack[l]{{\rm
${_{\l_6}}$}}}}}
  \end{picture}
\caption{An example of a 6-point spin-spin correlation function.}
\end{center}
\end{figure}
In the picture  we have that: $n=3$, $N=7$,
$a_1=1$, $a_2=2$, $a_3=7$, $a_4=10$, $a_5=13$, and $a_6=14$. The set
${\cal W}$ of
complex variables $w_j$ is: ${w_1, w_2, w_7, w_8, w_9, w_{10},
 w_{13}, w_{14}}$.

The 2n-CF $G_{2n}$ can now be written as:
\begin{equation}
  \label{eq:product2n}
 G_{2n} = (-i)^{\frac{L}{2}}
\left\{\prod_{w_j\in{\cal W}}\oint_{w_j}\right\}
{\cal G}\{w_j;\ze_i\}\times{\cal F}\{w_j\},
\end{equation}
where ${\cal F}\{w_j\}$ is a 'temperature' state:
\begin{equation}
  \label{eq:temperature}
{\cal F}\{w_j\} :=
\frac{\tr_{\Hn}\left(x^{2D^\rN}\prod_{w_j\in{\cal W}}
\phi^\rN(w_j)\right)}
{\tr_{\Hn}\left(x^{2D^\rN}\right)},
\end{equation}
and is known, see Sect.~\ref{sec:general}, to be a sum of products 
of Dirac-delta functions.
 
The function ${\cal G}\{w_j;\ze_i\}$ is given by the product of
$n$ blocks ${\sf  g}_i$, where
\begin{equation}
  \label{eq:g_i}
 {\sf g}_i := \!\!\!\!
\prod\limits_{a_{2i-1}\leq r <a_{2i}}\hspace{-0.6cm}{}^{'}
\;[w_rr]_\rR
\prod\limits_{a_{2i-1}<s \leq a_{2i}}\hspace{-0.6cm}{}^{'}
\;[w_ss]_\rN
\times\!\!\!
\prod\limits_{a_{2i-1}\leq k\leq a_{2i}}\prod\limits_{k<j\leq 2N}
[w_{k}j],
\end{equation}
in the above expression the parameters $r$, $s$, $k$, and $j$ are
positive integers.
The prime in the products means that we increment the parameters
$r$ and $s$ in steps of 2, i.e.~the $r$
runs over odd and the $s$ even integers respectively.

In the example presented before we have three blocks ${\sf  g}_1$,
${\sf  g}_2$, and ${\sf  g}_3$:
\begin{eqnarray*}
{\sf  g}_1 &=& [w_11]_\rR[w_22]_\rN
\prod_{j=1}^{2}\prod_{j<k}^{14}[w_jk]\\
{\sf  g}_2 &=& [w_77]_\rR[w_88]_\rN[w_99]_\rR[w_{10}10]_\rN
 \prod_{j=7}^{10}\prod_{j<k}^{14}[w_jk]\\ \\
{\sf  g}_3 &=&  [w_{13}13]_\rR[w_{14}14]_\rN[w_{13}14].             
\end{eqnarray*}
And ${\cal G} ={\sf  g}_1{\sf  g}_2{\sf  g}_3$.

The next step in the calculation is easy, namely to integrate the delta
functions in eq.~(\ref{eq:product2n}). The number of integrals is
reduced by a half and the final result is a clumsy sum of integrals
expressions. Perhaps some symmetrization procedure can give
more adequate results as we will show in the next section for the
two-point functions.

\section{Explicit results for the even correlation functions }
\label{sec:explicit}

In this section we review the general integration of the two-point
functions announced in~\cite{JMR}, details of the calculation are
relegated to the appendix. Also some examples of higher homogeneous 
nearest neighbors diagonal  CF are given in the form of determinants,
which generalize the well-known results for the diagonal 
two-point functions. Finally we treat the diagonal homogeneous four-point
functions for arbitrary spins locations.

\subsection{The general two-point functions}
\label{sec:general}

To integrate the expression found in Sect.~\ref{sec:arbitrary} we need 
the trace of the fermions operators in~(\ref{eq:product}) that
is given by a  a generalization of the lemma used before, i.e.
\begin{eqnarray*}
&&{\tr_{\Hr}\left(x^{2D^\rR}\phi^\rR(z_1)\phi^\rR(z_2)\cdots
\phi^\rR(z_{2n-1})\phi^\rR(z_{2n})\right)
\over \tr_{\Hr}\bigl(x^{2D^\rR}\bigr)}=\! 
\sum\limits_{p}{}^{'}
\!\varepsilon_p\delta^\rR(p_1p_2)\delta^\rR(p_3p_4)\cdots
\delta^\rR(p_{_{2n-1}}p_{_{2n}})\!,
\end{eqnarray*}
with the notation $\delta^\rR(ij)=\delta^\rR(x^2w_i/w_j)$. 
Here $p_1,\ldots, p_{2n}$ is some permutation of the numbers
$1,2,\ldots ,2n$, $\sum_p'$ 
is a summation over all $(2n-1)!!$ permutations which
satisfy the restrictions 
$p_{2m-1} < p_{2m}$ for $1<m<n$ and $p_{2m-1}
< p_{2m+1}$ for $1<m<n-1$ and 
$\varepsilon_p$ is  the parity of the permutation.
There is a similar expression 
for the $\rN$ case (note that the integrations 
in this case are well defined although the function $\delta^\rN(ij)$
has half-integral powers in the integration variable $z$) . 

Now changing the spectral parameters to their additive version~i.e.
interchanging $\ze_i$ with $\a(i)$ (remember that  
$\ze_i =\exp(-\pi\a(i)/2I)$) and using some identities for the product
of theta functions (see the appendix for the technicalities) we
arrive at the following simple expressions for the general two-point
functions 
\begin{eqnarray}
  \label{eq:two}
&&g^{\mu,\l}_{2n}=\frac{{\rm Pfaffian}
\left(h_{\mu,\l}(\a(i)-\a(j))\right)}
  {{\rm Pfaffian}\left(\sqrt{k}{\rm sn}(i\a(i)-i\a(j))\right)},\\
\label{eq:hlm}
&&h_\mu(\alpha):= 
-\frac{2I}{\pi}\left(\frac{k'}{k}\right)^{\frac{1}{2}}
\frac{\t1'(i\alpha)}{\th(i\alpha)},\quad
h_\l(\alpha):=
-\frac{2I}{\pi}(k')^{\frac{1}{2}}
\frac{\h1'(i\alpha)}{\th(i\alpha)}.
\end{eqnarray}
Expression (\ref{eq:two}) is symmetric in all spectral parameters
and satisfies the recurrence relation of Baxter\footnote{The integral
formula conjectured by Baxter~\cite{spin} for the general two-point function 
can also be derived see~\cite{JMR}.}
\begin{eqnarray*}
&&g^{\mu,\l}_{2n}(\a(1),\a(2),\cdots,\a(2n-1),\a(2n-1)+ I')= 
g^{\mu,\l}_{2n-2}(\a(1),\a(2),\cdots,\a(2n-2)),\n
&&g^{\mu,\l}_2(\a(1),\a(1)+I')=1.
\end{eqnarray*}
To obtain the homogeneous correlations one has to take the limit
of the expression (\ref{eq:two}) when all spectral parameters
$\a(i)$ are equal which requires  a 
generalization of the L'H\^opital's rule
for determinants already well known (see e.g. \cite{AP} or 
\cite{Yamada}), 
\begin{eqnarray}
\label{homogeneous}
&&g^{\mu,\l}_{hom,2n}=\frac{\det\left\{h_{\mu,\l}^{(2j+2k-3)}(0)
\right\} }
{\det\left\{h^{(2j+2k-3)}(0)\right\}}.
\end{eqnarray}
In the last formula $h^{n}(0)$ means the $n$'derivative of the
function $h$ evaluated at the point $0$,
$h(\alpha)= \sqrt{k}{\rm sn}(\alpha)$  and
$h_{\mu,\l}$ were already defined in (\ref{eq:hlm}). The determinants
have rank $n$ and their elements at the row $j$ and column $k$
are given by $h_{\mu,\l}^{(2j+2k-3)}(0)$. In the mathematical
literature they are called persymmetric Wronskians (a terminology
due to Sylvester). Because
the denominator in (\ref{homogeneous}) can be expressed as a product
one can calculate explicitly its homogeneous limit:
\begin{equation}
\label{eq:limit}
\det\left\{{\rm sn}^{(2j+2k-3)}(0)\right\}
=k^{n^2-n}\prod_{i=1}^{2n-1}(i!),
\end{equation}
therefore\footnote{Yamada has obtained a similar expression by
  different methods in~\cite{Yamada}.}
\begin{equation}
\label{eq:hom}
g^{\mu,\l}_{hom,2n}= k^{n-n^2}
{\prod_{i=1}^{2n-1}(i!)^{-1}}\times
\det\left\{\frac{1}{\sqrt{k}}h_{\mu,\l}^{(2j+2k-3)}(0)\right\},
\end{equation}
This expressions are useful for studying correlations for small
separations, as one only has to take derivatives 
of known elliptic functions. Several checks of eq.~(\ref{eq:hom}) with
the results in~\cite{GS} were done.

\subsection{Homogeneous even-point functions for adjacent spins on a
  diagonal }
\label{sec:even}

By adjacent spins I mean nearest diagonal neighbors. I discuss
the 4-point spin function and then the 6-point spin function the
generalization afterward will be  easy.

In the the case of the four-point CF ${\cal G}$ is the product of two blocks
${\sf g}_1$ and ${\sf g}_2$ and the temperature state ${\cal F}\{w_i\}$
is
\[
{\cal F}\{w_i\} = \delta^\rN\left(x^2 \frac{w_1}{w_2}\right)
\delta^\rN\left(x^2 \frac{w_5}{w_6}\right)-
\delta^\rN\left(x^2 \frac{w_1}{w_5}\right)
\delta^\rN\left(x^2 \frac{w_2}{w_6}\right)+
\delta^\rN\left(x^2 \frac{w_1}{w_6}\right)
\delta^\rN\left(x^2 \frac{w_2}{w_5}\right)
\]
The functions ${\sf g}_1$ and ${\sf g}_2$ are
\begin{eqnarray*}
 {\sf g}_1 = [w_11]_\rR[w_12][w_13][w_14] [w_15][w_16]&&\\
                  {} [w_22]_\rN[w_23][w_24] [w_25][w_26]&&
\end{eqnarray*}
\begin{eqnarray*}
{\sf g}_2 = [w_55]_\rR[w_56]&&\\
                {} [w_66]_\rN &&
\end{eqnarray*}
Doing the integration for the variables half of the variables $w_i$ in 
(\ref{eq:product2n}) gives
\begin{eqnarray}
  \label{app:4-point}
&&G_4 =  \oint_{w_1}[w_11]_\rR[w_22]_\rR\times
\oint_{w_5}[w_55]_\rR[w_56]_\rR - \n
&& -\oint_{w_1}[w_11]_\rR [w_12][w_13]
[w_14][w_15]_\rN\times\oint_{w_2}[w_22]_\rN[w_23][w_24][w_25]
[w_26]_\rR +\n
&&+\oint_{w_1}[w_11]_\rR[w_12][w_13][w_14][w_15][w_16]_\rR\times
\oint_{w_2}[w_22]_\rN[w_23][w_24][w_25]_\rN.
\end{eqnarray}
Equation (\ref{app:4-point}) is the inhomogeneous 4-point
function. Now I restrict to the homogeneous 4-point function i.e.,
take all spectral parameters $\ze_i$ equal to 1.

Let us change eq.~(\ref{app:4-point}) 
to the additive variables $v_i$ 
($w_i=\exp(i\pi v_i/I)$), but first I introduce the following 
notations:
\begin{equation}
\label{eq:coeff}
a_n := \frac{k^n}{\pi}\int_{-I}^{I}\!dv\,(\dn v)^2(\sn v)^{2n}
\qquad\mbox{for}\qquad n \in \Z. 
\end{equation}
Note that for $n<0$, $a_n$ can be put in a different form: by  the
shifting $v\rightarrow v +iI'$ in such a way that the poles of 
integrand  remain outside the contour of
integration and because the integrand is an even function
the limits of integration can be leave as in~(\ref{eq:coeff}), namely
\[
a_{-n} := -\frac{k^n}{\pi}\int_{-I}^{I}\!dv\,(\cn v)^2(\sn v)^{2(n-1)}
\qquad\mbox{for}\qquad n= 1,2,\ldots
\]
Returning to the homogeneous  4-point CF eq.~(\ref{app:4-point}), 
inserting the definitions for the different factors in the additive
variables
\begin{eqnarray*}
G_4 &= &
\left(\frac{1}{\pi}\int\!dv\, (\dn v)^2\right)^2 -
\left(\frac{1}{\pi}\int\!dv\, k^2\dn v\,\cn v\,
(\sn v)^3\right)^2 +\\
&+&\left(\frac{1}{\pi}\int\!dv\,k^2(\dn v)^2(\sn v)^4\right)\times
\left(\frac{1}{\pi}\int\!dv\,k^2(\cn v)^2(\sn v)^2\right),
\end{eqnarray*}
where the integrals go from $-I$ to $I$, 
the second term vanishes and with the above notation we have
\begin{equation}
  \label{app:4-point2}
 G_4 = a_0^2 - a_2\,a_{-2}. 
\end{equation}
Proceeding in the same way  the 6-point CF is given by
\begin{equation}
  \label{app:6-point}
 G_6 = a_0^3 -2a_0\,a_2\,a_{-2} +a_2^2\,a_{-4} +a_4\,a_{-2}^2-
a_0\,a_4\,a_{-4} =\left|\,\matrix{
a_0\hfill& a_{-2}\hfill& a_{-4}\cr
a_2\hfill& a_0\hfill& a_{-2}\cr
a_4\hfill& a_2\hfill& a_0\cr}\right|.
\end{equation}
The general expression for $G_{2n}$ can be induced to be:
\begin{equation}
  \label{eq:2n-point}
G_{2n} = \left|\,\matrix{
a_0\hfill& a_{-2}\hfill\cdots&a_{-2(n-1)}\cr
a_2\hfill& a_0\hfill\cdots&a_{-2(n-2)}\cr
\,\vdots\hfill&\,\vdots\hfill&\,\vdots\hfill\cr
a_{2(n-1)}\hfill& a_{2(n-2)}\hfill\cdots&a_0\cr}\right|.
\end{equation}

\Remark 
The definitions of the coefficients of the above
matrices is equivalent to those given on~\cite{MT} pag.~199.
For temperatures below
the critical temperature, when $0< k<1$  they are 
given by
\begin{equation}
  \label{remark-coeff}
a_n = \frac{1}{2\pi}
\int_{0}^{2\pi}d\theta e^{-in\theta}
\left[\frac{k^{-1}-e^{-i \theta}}{k^{-1}
-e^{i \theta}}\right]^\half,
\end{equation}
deforming the contour of integration from the unit circle $|\xi| =1$
($\xi := e^{-i\theta}$) to the real axis one  rewrites
(\ref{remark-coeff}) as
\[
a_n = \frac{1}{\pi}
\int_{0}^{k}\frac{d\xi}{\xi}\xi^n\left(\frac{1-k\xi}{k\xi^{-1}-1}
\right)^\half,
\]
substituting $\xi = k(\sn u)^2$ one sees that the definition 
(\ref{remark-coeff}) coincides with (\ref{eq:coeff}).

In a similar manner the Toeplitz determinant for the two-point
function (see e.g. \cite{MT}) can be derived by the method presented
above, giving a strong support of the vertex operator method.

\subsection{Homogeneous diagonal Four-point functions}
\label{sec:diagonal}

We label the four spins from left to right as $\l_1$, $\l_2$,
$\l_3$, $\l_4$ and let the distance between the $\l_1$ and
$\l_2$, be $r$ and  between $\l_3$ and  $\l_4$ be $s$, finally
let the distance between $\l_2$ and $\l_3$ be $p$, where the distances
are measured in diagonal steps of course. Then by the same procedure 
as before, we have the following general result.

The CF can be written as a determinant $G$ of order $r+s$ whose 
elements on the principal diagonal are all equal to $a_0$ and
whose elements $a_n$ above and below the diagonal are negative
and positive sub-indices 
respectively. To give its form decompose $G$ in four matrix blocks: 
\begin{equation}
  \label{eq:blocks}
  G = \left|\matrix{A_1\hfill & B_1\cr
                      B_2\hfill & A_2\cr}\right|.
\end{equation}
$A_1$ is the same Toeplitz matrix of order $r\times r$
for the two-point function
$\<\l_1\l_2\>$ i.e.
\[
A_1 := \left(\,\matrix{
a_0\hfill& a_{-1}\hfill\cdots&a_{-r+1}\cr
a_1\hfill &a_0\hfill\cdots&a_{-r+2}\cr
\,\vdots\hfill&\,\vdots\hfill&\,\vdots\hfill\cr
a_{r-1}&a_{r-2}\cdots&a_{0}\cr}\right).
\]
$A_2$ is also a Toeplitz matrix but now of order $s$.
$B_1$ is a $r\times s$ matrix with all its $a_n$ coefficients
negative whose element in the first row and at the $s$ column
is equal to $a_{-(r+s+p)+1}$ and with coefficient sub-indices 
growing to the left and to the bottom of the matrix
\[
B_1 := \left(\,\matrix{
a_{-(r+p)}\hfill& a_{-(r+p)-1}\hfill\cdots&a_{-(r+s+p)+1}\cr
a_{-(r+p)+1}\hfill &a_{-(r+p)}\hfill\cdots&a_{-(r+s+p)+2}\cr
\,\vdots\hfill&\,\vdots\hfill&\,\vdots\hfill\cr
a_{-p-1}&a_{-p-2}\cdots&a_{-(s+p)}\cr}\right).
\]
Finally the matrix $B_2$ is the transpose of $B_1$ but with all its
coefficients positive.

Let us give some examples for illustration. Consider the following
configuration where each circle represents a point in the diagonal:
\begin{center}
 \setlength{\unitlength}{0.0125in}
\begin{picture}(150,23)
\put(10,10){\makebox(0,0)[lb]{\raisebox{0pt}[0pt][0pt]{\shortstack[l]{{\rm
$\bu$}}}}}
\put(30,10){\makebox(0,0)[lb]{\raisebox{0pt}[0pt][0pt]{\shortstack[l]{{\rm
$\circ$}}}}}
\put(50,10){\makebox(0,0)[lb]{\raisebox{0pt}[0pt][0pt]{\shortstack[l]{{\rm
$\bu$}}}}}
\put(70,10){\makebox(0,0)[lb]{\raisebox{0pt}[0pt][0pt]{\shortstack[l]{{\rm
$\circ$}}}}}
\put(90,10){\makebox(0,0)[lb]{\raisebox{0pt}[0pt][0pt]{\shortstack[l]{{\rm
$\circ$}}}}}
\put(110,10){\makebox(0,0)[lb]{\raisebox{0pt}[0pt][0pt]{\shortstack[l]{{\rm
$\bu$}}}}}
\put(130,10){\makebox(0,0)[lb]{\raisebox{0pt}[0pt][0pt]{\shortstack[l]{{\rm
$\bu$}}}}}
\put(8,2){\makebox(0,0)[lb]{\raisebox{0pt}[0pt][0pt]{\shortstack[l]{{\rm
$\l_1$}}}}}
\put(48,2){\makebox(0,0)[lb]{\raisebox{0pt}[0pt][0pt]{\shortstack[l]{{\rm
$\l_2$}}}}}
\put(108,2){\makebox(0,0)[lb]{\raisebox{0pt}[0pt][0pt]{\shortstack[l]{{\rm
$\l_3$}}}}}
\put(128,2){\makebox(0,0)[lb]{\raisebox{0pt}[0pt][0pt]{\shortstack[l]{{\rm
$\l_4$}}}}}
\end{picture}
\end{center}
The correlation function for this configuration is (with $r=2$, $p=3$ and
$s=1$):
\[ 
\left|\,\matrix{a_0\hfill& a_{-1}\hfill& a_{-5}\cr
           a_1\hfill& a_{0}\hfill& a_{-4}\cr
           a_5\hfill& a_{4}\hfill& a_{0}\cr}\right|.
\]

An interesting example is given by the following picture
\begin{center}
 \setlength{\unitlength}{0.0125in}
\begin{picture}(200,30)
\put(10,10){\makebox(0,0)[lb]{\raisebox{0pt}[0pt][0pt]{\shortstack[l]{{\rm
$\bu$}}}}}
\put(30,10){\makebox(0,0)[lb]{\raisebox{0pt}[0pt][0pt]{\shortstack[l]{{\rm
$\bu$}}}}}
\put(50,10){\makebox(0,0)[lb]{\raisebox{0pt}[0pt][0pt]{\shortstack[l]{{\rm
$\circ$}}}}}
\put(70,10){\makebox(0,0)[lb]{\raisebox{0pt}[0pt][0pt]{\shortstack[l]{{\rm
$\circ$}}}}}
\put(90,10){\makebox(0,0)[lb]{\raisebox{0pt}[0pt][0pt]{\shortstack[l]{{\rm
$\;\cdots$}}}}}
\put(130,10){\makebox(0,0)[lb]{\raisebox{0pt}[0pt][0pt]{\shortstack[l]{{\rm
$\circ$}}}}}
\put(150,10){\makebox(0,0)[lb]{\raisebox{0pt}[0pt][0pt]{\shortstack[l]{{\rm
$\circ$}}}}}
\put(170,10){\makebox(0,0)[lb]{\raisebox{0pt}[0pt][0pt]{\shortstack[l]{{\rm
$\bu$}}}}}
\put(190,10){\makebox(0,0)[lb]{\raisebox{0pt}[0pt][0pt]{\shortstack[l]{{\rm
$\bu$}}}}}
\put(8,2){\makebox(0,0)[lb]{\raisebox{0pt}[0pt][0pt]{\shortstack[l]{{\rm
$\l_1$}}}}}
\put(28,2){\makebox(0,0)[lb]{\raisebox{0pt}[0pt][0pt]{\shortstack[l]{{\rm
$\l_2$}}}}}
\put(168,2){\makebox(0,0)[lb]{\raisebox{0pt}[0pt][0pt]{\shortstack[l]{{\rm
$\l_3$}}}}}
\put(188,2){\makebox(0,0)[lb]{\raisebox{0pt}[0pt][0pt]{\shortstack[l]{{\rm
$\l_4$}}}}}
\end{picture}
\end{center}
in the figure we suppose that the distance between $\l_2$ and $\l_3$
is $n$ (i.e. $p =n$). Then the CF is
\[
\left|\matrix{a_0\hfill & a_{-n-1}\cr
                      a_{n+1}\hfill & a_0\cr}\right|,
\]
these coincides with the value found 
in~\cite{S} using combinatorial methods.

\section{Odd correlation functions}
\label{sec:odd-point}

In this section we want to give some examples of odd-point spin CF
(there are not odd disorder CF under the critical temperature).
As example of higher odd-point functions I treat some particular cases
of the three spin CF .

\begin{figure}[ht]
\begin{center}
 \setlength{\unitlength}{0.0125in}
  \begin{picture}(300,60)(100,0)
\drawline(200,30)(220,50)(240,30)(260,50)(280,30)
(260,10)(240,30)(220,10)(200,30)
\dashline{2}(210,5)(210,55)
\dashline{2}(230,5)(230,55)
\dashline{2}(250,5)(250,55)
\dashline{2}(270,5)(270,55)
\put(207,0){\makebox(0,0)[lb]{\raisebox{0pt}[0pt][0pt]{\shortstack[l]{{\rm
${_{\ze_{1}}}$}}}}}
\put(227,0){\makebox(0,0)[lb]{\raisebox{0pt}[0pt][0pt]{\shortstack[l]{{\rm
${_{\ze_{2}}}$}}}}}
\put(247,0){\makebox(0,0)[lb]{\raisebox{0pt}[0pt][0pt]{\shortstack[l]{{\rm
${_{\ze_{3}}}$}}}}}
\put(267,0){\makebox(0,0)[lb]{\raisebox{0pt}[0pt][0pt]{\shortstack[l]{{\rm
${_{\ze_{4}}}$}}}}}
\put(195,23){\makebox(0,0)[lb]{\raisebox{0pt}[0pt][0pt]{\shortstack[l]{{\rm
${_{\l_1}}$}}}}}
\put(235,23){\makebox(0,0)[lb]{\raisebox{0pt}[0pt][0pt]{\shortstack[l]{{\rm
${_{\l_2}}$}}}}}
\put(275,23){\makebox(0,0)[lb]{\raisebox{0pt}[0pt][0pt]{\shortstack[l]{{\rm
${_{\l_3}}$}}}}}
  \end{picture}
\caption{Adjacent 3-point  correlation function.}
 \label{fig:3}
\end{center}
\end{figure}
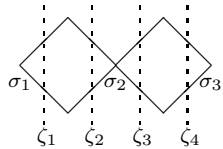

Consider first the three spins in adjacent diagonal positions see
Fig.~\ref{fig:3}.
Let us call the 3-point CF $C_1(\a(1),\a(2),\a(3),\a(4))$ (the
chose of the index will be clear below). Its
explicit expression is given by
\begin{eqnarray}
  \label{eq:3-point}
&&C_1 = \frac{1}{\cal N}\sum\limits_{\l_i}\l_1\l_2\l_3\times\n
&&\quad\tr_{\Hn}
\left(x^{2D^\rN}\Pu(\l_3,x\ze_4)\Pd(\l_2,x\ze_3)\Pu(\l_2,x\ze_2)
 \Pd(\l_1,x\ze_1)\Pu(\l_1,\ze_1)\Pd(\l_2,\ze_2)\Pu(\l_2,\ze_3)
 \Pd(\l_3,\ze_4)\right), 
\end{eqnarray}
where the normalization factor 
${\cal N}$  is  $(g^\rR g^\rN)^2\tr_\Hn(x^{2D^\rN})$.
Now using eqs.~(\ref{eq:inter1}) and (\ref{eq:inter1.2}) and
starting at the middle of the operator expression
in~(\ref{eq:3-point}). After using the unitarity conditions for the
VOs we arrive at
\[
\sum\limits_{\l_3}\l_3\Pu(\l_3,x\ze_4)\Pd(\l_3,\ze_4) =
g^\rN(\id_{\H^{\rN,+}}-\id_{\H^{\rN,-}}) =g^\rN\psi_0^\rN,
\]
where the unitarity relation~(\ref{eq:inverse2})
 was employed and the fermion operator
$\psi_0^\rN$  anticommutes with all the fermions in the $\rN$
representation and has eigenvalue $+1$ in even subspace and $-1$ at
the odd subspace. Collecting all the terms we have
\begin{eqnarray}
 \label{eq:3-point2}
 C_1 = -i\oint_{w_1}\oint_{w_2} {\sf g}(w_1,w_2)\times
 \mathcal{F}(w_1,w_2),
\end{eqnarray}
\begin{eqnarray*}
{\sf g}(w_1,w_2) = [w_11]_\rR[w_12][w_13][w_14] &&\\
                  {} [w_22]_\rN[w_23][w_24],&&
\end{eqnarray*}
and
\begin{equation}
 \label{eq:trace-odd}   
 \mathcal{F}(w_1,w_2) = 
\frac{\tr_{\Hn}\left(x^{2D^\rN}\phi^\rN(w_1)\phi^\rN(w_2)\psi_0^\rN\right)}
{\tr_{\Hn}\left(x^{2D^\rN}\right)}.
\end{equation}            
Evaluating the expression~(\ref{eq:trace-odd}) gives 
\begin{equation}
  \label{eq:trace-eval}
 \mathcal{F}(w_1,w_2) = M\left(\delta^\rN(x^2\frac{w_1}{w_2})
+ \frac{2kI}{i\pi}\sn (v_1-v_2)\right), 
\end{equation}
$M$ being the magnetization\footnote{The operator
$\psi_0^\rN$  represents  the central spin matrix $\l_1^z$ under the
Jordan-Wigner transformation, 
see~\cite{FJMMN}.}
\[
M := \frac{tr_{\Hn}\left(x^{2D^\rN}\psi_0^\rN\right)}
{tr_{\Hn}\left(x^{2D^\rN}\right)}     =(1-k^2)^{1/8}.
\]
Carrying out  the integrations explicitly the three-point function is
\begin{eqnarray}
  \label{eq:3-point3}
C_1 = &&M\prod\limits_{1\leq i< j\leq 4}
\left({\rm sn}(i\a(ij))\right)^{-1}\times\n
&&\Big\{\dn(i\a(24))\sn(i\a(14))[\sn(i\a(23))-\cn(i\a(12))\sn(i\a(13))
+\sn(i\a(12))\cn(i\a(13))]\n
&&-\dn(i\a(23))\sn(i\a(13))[\sn(i\a(24))-\cn(i\a(12))\sn(i\a(14))
+\sn(i\a(12))\cn(i\a(14))]\\
&&\qquad+\sn(i\a(12))[\sn(i\a(34))-\cn(i\a(13))\sn(i\a(14))
+\sn(i\a(13))\cn(i\a(14))]\Big\}\nonumber.
\end{eqnarray}
To make explicit the symmetries of this function, we employ the 
identities, $\cn(u-v)=\cn u\cn v+\sn u\sn v\dn(u-v)$ and
$\dn u\sn(u-v)= -\cn u\sn v +\sn u\cn v\dn(u-v)$, obtaining
\begin{eqnarray}
  \label{eq:3-point-sim}
&&C_1 = M\!\!\!\prod\limits_{1\leq i< j\leq 4}
\left({\rm sn}(i\a(ij))\right)^{-1}
\Big\{\sn(i\a(12))\sn(i\a(34))+\cn(i\a(13))\cn(i\a(24))
-\cn(i\a(14))\cn(i\a(23))+\n
&&\sn(i\a(14))\sn(i\a(23))[\dn(i\a(13))+
\dn(i\a(24))]-\sn(i\a(13))\sn(i\a(24))[\dn(i\a(14))+\dn(i\a(23))]\Big\},
\end{eqnarray}
eq.~(\ref{eq:3-point-sim}) has now symmetries under the transpositions
$1\lr 2$ or $3\lr 4$ and the geometrical symmetry under the double
transposition $1\lr4$ and $2\lr 3$ see Fig.~\ref{fig:3};
note that the above expression is \emph{not}
invariant under the sole transposition $2\lr 3$ or $1\lr 4$.

\begin{figure}[ht]
  \begin{center}
    \setlength{\unitlength}{0.0125in}
\begin{picture}(200,30)
\put(10,10){\makebox(0,0)[lb]{\raisebox{0pt}[0pt][0pt]{\shortstack[l]{{\rm
$\bu$}}}}}
\put(30,10){\makebox(0,0)[lb]{\raisebox{0pt}[0pt][0pt]{\shortstack[l]{{\rm
$\bu$}}}}}
\put(50,10){\makebox(0,0)[lb]{\raisebox{0pt}[0pt][0pt]{\shortstack[l]{{\rm
$\circ$}}}}}
\put(70,10){\makebox(0,0)[lb]{\raisebox{0pt}[0pt][0pt]{\shortstack[l]{{\rm
$\circ$}}}}}
\put(90,10){\makebox(0,0)[lb]{\raisebox{0pt}[0pt][0pt]{\shortstack[l]{{\rm
$\;\cdots$}}}}}
\put(130,10){\makebox(0,0)[lb]{\raisebox{0pt}[0pt][0pt]{\shortstack[l]{{\rm
$\circ$}}}}}
\put(150,10){\makebox(0,0)[lb]{\raisebox{0pt}[0pt][0pt]{\shortstack[l]{{\rm
$\circ$}}}}}
\put(170,10){\makebox(0,0)[lb]{\raisebox{0pt}[0pt][0pt]{\shortstack[l]{{\rm
$\circ$}}}}}
\put(190,10){\makebox(0,0)[lb]{\raisebox{0pt}[0pt][0pt]{\shortstack[l]{{\rm
$\bu$}}}}}
\put(8,2){\makebox(0,0)[lb]{\raisebox{0pt}[0pt][0pt]{\shortstack[l]{{\rm
$\l_1$}}}}}
\put(28,2){\makebox(0,0)[lb]{\raisebox{0pt}[0pt][0pt]{\shortstack[l]{{\rm
$\l_2$}}}}}
\put(188,2){\makebox(0,0)[lb]{\raisebox{0pt}[0pt][0pt]{\shortstack[l]{{\rm
$\l_3$}}}}}
\end{picture}
\caption{3-point function with one spin at arbitrary distance.}
    \label{fig:3n}
  \end{center}
\end{figure}
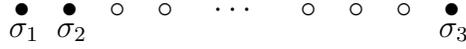

Taking all the $\a(i)$ in (\ref{eq:3-point3})or (\ref{eq:3-point-sim})
equal and using (\ref{eq:limit}) we obtain 
the homogeneous diagonal three-point function\footnote{This limit 
can also be done, of course, in the integral expression 
first and then integrate by residues.}
\begin{equation}
  \label{eq:homo-3}
  C_1^{hom} =M\left(1 -\frac{k^2}{4}\right).
\end{equation}
Let us investigate the case that the third spin goes to infinity
the other spins remaining adjacent diagonal neighbors in the
homogeneous case Fig.~\ref{fig:3n}. By the cluster property we expect
that the CF tends to the product of spin expectation value times
the two-point function as the spin $\l_3$ tends to infinity,
this limit will be another check for the VOM.
The double integral expression for the homogeneous case is then
\begin{eqnarray}
  \label{eq:3-arb}
C_n^{hom}\!\!\! := M\frac{k^{2n+2}}{\pi^2}\!
\int_{-I}^{I}\!\!\int_{-I}^{I}\!\!\!dv_1dv_2
\dn v_1\sn^{2n+1}v_1\cn v_2\sn^{2n}v_2\left[
\sn(v_1\! -\! v_2)\!+\!\frac{i}{k}\delta^\rN(v_1\! -\! v_2+\!iI')\right],
\end{eqnarray}
where $n=1$ corresponds to the case of adjacent diagonal spins.
The integration of the delta function gives
\begin{equation}
  \label{eq:delta}
  \frac{1}{\pi}\int_{-I}^{I}\!\!dv\,\dn^2v = \frac{2}{\pi}E(k),
\end{equation}
where $E(k)$ is the  complete elliptic integral of second kind. 
The integration
of the first term in~(\ref{eq:3-arb}) could be done in principle by
residues but the most simple way seems to be to use the addition
formula for the function $\sn(v_1-v_2)$,
\[
\sn(v_1-v_2) = \frac{\sn v_1\cn v_2\dn v_2-\sn v_2\cn v_1\dn v_1}
{1 - k^2\sn^2 v_1\sn^2 v_2},
\]
substituting this formula we see that only the first factor is no zero
on symmetry grounds
\begin{eqnarray}
 \label{eq:I_n}
  I_n = \frac{k^{2n+2}}{\pi^2}\int_{-I}^{I}\!\!\int_{-I}^{I}
\!\! dv_1dv_2\,\frac{\sn^{2n+2} v_1\sn^{2n} v_2\cn^2 v_2\dn v_1\dn v_2}
{1 - k^2\sn^2 v_1\sn^2 v_2}.
\end{eqnarray}
Define the following integrals
\begin{equation}
  \label{eq:J_n}
  J_n := \int_{-I}^{I}\!\!dv\,\dn v\,\sn^{2n}v,
\end{equation}
by differentiation of the expression $\cn v\,\sn^{2n}v$ it can be proved
that the integrals $J_n$ satisfy the following recursion relations
\[
J_{n+1} = \frac{2n+1}{2n+2}J_n.
\]
In the interval $-I\leq v_1,v_2\leq I$ we can  expand the 
denominator in~(\ref{eq:I_n})
\[
(1 - k^2\sn^2 v_1\sn^2 v_2)^{-1} =\sum\limits_{j=0}^{\infty}
k^{2j}\sn^{2j} v_1\sn^{2j} v_2,
\]
putting this expression under the integrations an using the
definitions of $J_n$ one arrives to the following series (interchange
of the series with the integrals is allowed due to the uniform
convergence of the series in the interval of integration)
\[
I_n = \frac{k^{2n+2}}{\pi^2}\left[\sum\limits_{j=0}^{\infty}
k^{2j}J_{n+j+1}\left(J_{n+j}-J_{n+j+1}\right)\right],
\]
using the recursion relations and changing the index of summation 
one has
\[
I_n =\frac{1}{\pi^2}\sum\limits_{j=n+1}^{\infty}\frac{k^{2j}J_{j}^2}
{2j -1},
\]
now because $J_0 =\pi$ by the recursion relations
\[
J_{j} = \frac{(2j-1)!!}{(2j)!!}\pi,
\]
finally
\[
I_n =\sum\limits_{j=n+1}^{\infty}k^{2j}
\left[\frac{(2j-1)!!}{(2j)!!}\right]^2\frac{1}{2j -1},
\]
by the series definition of the elliptic integral $E(k)$ 
(see e.g.~\cite{grad})
\[
I_0 = 1 - \frac{2}{\pi}E(k)
\]
we also have that
\[
I_n -I_0 = -\sum\limits_{j=1}^{n}k^{2j}
\left[\frac{(2j-1)!!}{(2j)!!}\right]^2\frac{1}{2j -1},
\]
therefore the final expression for the 3-point function $G_n$ is
\begin{equation}
  \label{eq:infty}
  C_n^{hom} = M\left\{1 -\sum\limits_{j=1}^{n}k^{2j}
\left[\frac{(2j-1)!!}{(2j)!!}\right]^2\frac{1}{2j -1}\right\},
\end{equation}
with $n\rightarrow\infty$ $G_n$ tends to $M\times\frac{2}{\pi}E(k)$
i.e. the product of the magnetization with the two-point function with 
as we expect.

\section{Conclusion}
\label{sec:con}

We have seen in the calculation of different CF 
how intimate is the connection of the IM with the theory 
of elliptic functions and how the vertex operator method allows 
to make this connection very directly. In this way several scattered 
results about the CF of the Ising model in the literature
could be treated in a general way. Because the method is not totally
rigorous, specially in the definition of the CTM, several consistency
checks were done: comparision with results already known by other methods
or physical reasonable consequences.

We have only treated purely
order or disorder CF but with the same methods exposed in this article
one can write immediately integral expressions for the mixed CF. A more
difficult problem is to obtain a complete integration of the higher CF
as was done for the two-point functions or alternatively to obtain
possible equations satisfied by the CF like the ones found
in~\cite{Mccoy} and references therein. Also the generalized Toeplitz 
determinants found in Sect.~\ref{sec:even} and \ref{sec:diagonal} could
be used to see what kind of nonlinear differential equations the
CF satisfy in the same way as was done in~\cite{Sato}.

\paragraph{Acknowledgments}
Supported in part by DFG Sfb 288 ``Differentialgeometrie und
Quantenphysik'' and by research funds of the Freie Universit\"at
Berlin.


\appendix
\section{Integration of the two-point correlation function}
\label{app:integration}

Here I give the details of the explicit integration of order-order
correlation function. The disorder-disorder case is analogous and
I will indicate in passing the modification needed. 

I begin remembering the expression obtained for the disorder-disorder
CF on page~\pageref{eq:product} equation (\ref{eq:product}), 
that I write know in the following way
\begin{equation}
  \label{app:prod}
  g_{2n}^\mu = \left\{\prod_{1\leq i \leq 2n}
\oint_{w_i}\right\} {\cal G}_{2n}^\rR
\{w_i\}\times{\cal F}_{2n}^\rR\{w_i\},
\end{equation}
where
\begin{equation}
 \label{app:prod,deltas}
{\cal F}_{2n}^\rR\{w_i\} := \sum\limits_p{}{}^{'} 
\ve_p\delta^\rR(p_1p_2)\delta^\rR(p_3p_4)\cdots
\delta^\rR(p_{2n-1}p_{2n}),
\end{equation}
with the same notation as in Sect.~\ref{sec:general}, and
\begin{equation}
 \label{app:prod,funct}
{\cal G}_{2n}^\rR\{w_i\} := (-i)^n
[w_11]_{\rN}[w_22]_{\rR}\cdots
[w_{2n-1}(2n-1)]_{\rN}[w_{2n}2n]_{\rR}\times\!\!\!\!\! 
\prod_{1\leq i < j \leq 2n}[w_ij].
\end{equation}
As I showed in~\cite{JMR}, 
before inserting this formula under the integral
expression it is more convenient to antisymmetrize the factor
multiplying the fermions in all the $w_i$ variables (under the
integrals the fermions can be anticommutated as already observed),
for this we change the spectral parameters and integration variables
to their additive version,
\begin{eqnarray*}
&&\oint_{w_i}\rightarrow \int_{-I}^{I}{dv_i\over 2I},\hspace{3cm}
\delta^{\rN,\rR}(ij)\rightarrow 2I
\delta^{\rN,\rR}(v_i-v_j+iI'), \\
&&[w_ii]_{\rN}=\sqrt{2Ik'\over\pi}{\h1(v_i-i\a(i))
\over\th(v_i-i\a(i))},
\quad[w_ii]_{\rR}=\sqrt{2Ik'\over\pi}{\t1(v_i-i\a(i))
\over\th(v_i-i\a(i))},\\
&&[w_ij]     = -{\hh(v_i-i\a(j))\over\th(v_i-i\a(j))},
\end{eqnarray*}
The result of the antisymmetrization is 
\begin{eqnarray}
\label{eq:theta}
&&{\prod_{i=1}^n\h1(v_{2i-1}-i\a(2i-1))\t1(v_{2i}-i\a(2i))
\prod_{1\leq i < j \leq 2n}\hh(v_i-i\a(j))\over 
\prod_{1\leq i \leq j \leq 2n}\th(v_i-i\a(j))}+ \mbox{perm} = \n
&&{\prod_{1\leq i < j \leq 2n}\th(i\a(i) - i\a(j))\hh(v_i-v_j)
\left(\h1(0)\t1(0)\right)^n \over \h1(0)\prod_{i=1}^{2n}
\prod_{j=1}^{2n}
\th(v_i-i\a(j))}\!\times\!
\h1\!\!\left(\sum_{i=1}^{2n}v_i-i\sum_{r=1}^{2n}
\a(r)\!\right),
\end{eqnarray}
where by perm we mean the antisymmetrization in all $v_i$. The same
formula is valid under the interchange of $\t1$ with $\h1$. This
identity is proved showing first that both sides transform
identically under the periods $2I$ and $2iI'$ in all $v_i$, and then
through an inductive process it is possible show that  
the residues of the simple poles
at e.g. $v_1=i\a(1)+iI'$ are identical for the two sides.
As a simple consequence of Liouville's theorem the rhs can only
differ by an additive constant, which is proved to be zero
by comparing the two sides at e.g. $v_{2n}=i\a(2n)$. Therefore we
can now write
\begin{equation}
  \label{app:G,trans}
{\cal G}_{2n}^\rR\{v_i\} =
\frac{c}{(2n)!}\times
\frac{\prod\limits_{1\leq i < j\leq 2n}\th(i\a(ij))\hh(v_{ij})}
{\h1(0)\prod\limits_{i,j=1}^{2n}\th(v_i-i\a(j))}
\!\times\! \h1\!\!\left(\sum_{i=1}^{2n}v_i-i\sum_{r=1}^{2n}
\a(r)\!\right),
\end{equation}
$c := (ik'\h1(0)\t1(0)/\pi)^n$ is a constant factor;  
$\a(ij)=\a(i)-\a(j)$ and $v_{ij}=v_i-v_j$.

Comparing the rhs. of the above equation with the rhs. 
of the following classical result\footnote{This formula is incorrectly
  given in reference~\cite{JMR}.} (see e.g. App.~A in~\cite{Yamada})
\begin{eqnarray*}
\det\limits_{1\leq i< j\leq 2n}\!\!
\left[\frac{\t1(v_i-i\a(j))\hh(0)'}{\th(v_i-i\a(j))\h1(0)}\right]=
\frac{\left(\hh'(0)\right)^{2n}}{\h1(0)}
\frac{\prod\limits_{i<j}\hh(v_{ij})
\hh(i\a(ij))}{\prod\limits_{i,j}\th(v_i-i\a(j))}
\h1\!\left(\!\sum_i(v_i-i\a(i))\!\right),
\end{eqnarray*}
we see that by multiplication of some factors  dependent on
the variables $v_i$ one can write
\begin{equation}
  \label{app:G2}
{\cal G}_{2n}^\rR\{v_i\} = B(\a(i))\times\frac{1}{(2n)!}\times
\det\limits_{1\leq i< j\leq 2n}\left[\hat{h}_\l(v_i-\a(j))\right]
\end{equation}
with
\begin{equation}
 \label{app:hat{h}}
\hat{h}_\mu(v) :=
\frac{\t1(v)\hh(0)'}{\th(v)\h1(0)}= \dn v.
\end{equation}
In the order case one has instead $\hat{h}_\l(v):= k\cn v$.

The function $B(\a(i))$ is:
\begin{equation}
 \label{app:B}
B(\a(i)) := \frac{c}{(\hh'(0))^{2n}}\prod\limits_{i<j}
\frac{\th(i\a(ij))}{\hh(i\a(ij))} = 
\left(\frac{i}{\pi \sqrt{k}}\right)^n
\left[\prod\limits_{i<j}\sqrt{k}\,\sn i\a(ij)\right]^{-1}.
\end{equation}

Inserting eqs.~(\ref{app:G2}),(\ref{app:hat{h}}) and 
(\ref{app:B}) under the integration symbols gives (I consider
now simultaneously the order case also)
\begin{eqnarray}
 \label{app:glmu}
&&g_{2n}^{\l,\mu}= B(\a(i))\times\left\{\prod\limits_{i=1}^{2n}
\int_{-I}^{I}dv_i\right\}\frac{1}{(2n)!}\sum\limits_{P}\varepsilon_P
\prod\limits_{j=1}^{2n}\hat{h}_{\l,\mu}(v_j-i\a({Pj}))\times\n
&&\sum\limits_Q{}^{'}\varepsilon_Q
\delta(v_{Q1}-v_{Q2}+iI')\cdots
\delta(v_{Q(2n-1)}-v_{Q2n}+iI').
\end{eqnarray}
Now each of the $(2n-1)!!$ terms in the sum over the delta functions
gives the same result when integrated over their respective
variables $v_i$, because both factors under the integrals are
totally antisymmetric in the $v_i$'s, therefore it is only necessary
to integrate over the first term in the second sum i.e., 
$\delta(v_1-v_2+iI')\delta(v_3-v_4+iI')\cdots
\delta(v_{2n-1}-v_{2n}+iI')$ multiplied by
the constant $(2n-1)!!/(2n)! =1/(2^n\, n!)$. Integrating out the
deltas one is left with (after renaming the integration variables)
\begin{eqnarray*}
&&g_{2n}^{\l,\mu}= B(\a(i))\times\left\{\prod\limits_{i=1}^{n}
\int_{-I}^{I}dv_i\right\}\frac{1}{2^n n!}
\prod\limits_{j=1}^{n}
\sum\limits_{P}\varepsilon_P\hat{h}_{\l,\mu}(v_j-i\a(Pj))
\hat{h}_{\l,\mu}(v_j-i\hat{\alpha}_{P(j+1)})=\\
&&\left[\prod\limits_{i<j}\sqrt{k}\,\sn i\a(ij)\right]^{-1}
\!\!\!\times\frac{1}{2^n n!}
\sum\limits_{P}\varepsilon_P h_{\l,\mu}(i\a(P1)-i\hat{\alpha}_{P2})
\cdots
h_{\l,\mu}(i\alpha_{P(2n-1)}-i\hat{\alpha}_{P2n}),
\end{eqnarray*}
where $\hat{\alpha}=\alpha -I'$ and the following definition was
introduced:
\begin{eqnarray*}
  h_{\l,\mu}(i\a(i)-i\a(j)) :=\frac{i}{\sqrt{k}\pi}\int_{-I}^{I}dv 
\hat{h}_{\l,\mu}(v-i\a(i))\hat{h}_{\l,\mu}(v-i\a(j)).
\end{eqnarray*}
For the integration of the last formula we use (see \cite{Yamada1})
\begin{equation}
 \int_{-I}^{I}du\,\cn (u+ia)\cn (u+ib)=
2I[\cn (ia-ib)+ik^{-1}
\cn (ia-ib-iI')\,\zn (ia-ib)]\label{app:cn} 
\end{equation}
\begin{equation}
\int_{-I}^{I}du\,\dn (u+ia)\dn (u+ib)=
2I[\dn (ia-ib)+i
\dn (ia-ib-iI')\,\zn (ia-ib)]\label{app:dn}. 
\end{equation}

Equations (\ref{app:cn}) and (\ref{app:dn}) can
be shown under the condition
\[
-I'<a,b<I'
\]
for real numbers $a$ and $b$. $\zn u = \th'(u)/\th(u)$ 
is the Jacobi zeta function. Employing these results we have by
analytic continuation,
\begin{eqnarray}
  \label{hsigma}
h_\l(i\a(i)-i\hat{\alpha}_{j}) &=& \frac{2I\sqrt{k}}{\pi}
[\dn i\a(ij)\,\sn i\a(ij) -\cn i\a(ij)\,\zn i\a(ij)] 
+ i\sqrt{k}\cn \a(ij) \\
\label{hmu}
h_\mu(i\a(i)-i\hat{\alpha}_{j}) &=& \frac{2I}{\sqrt{k}\pi}
[k^2\cn i\a(ij)\,\sn i\a(ij) -\dn i\a(ij)\,\zn i\a(ij)] 
+ \frac{i}{\sqrt{k}}\dn i\a(ij).
\end{eqnarray}
The  terms in square brackets in
eqs.~(\ref{hsigma}) and (\ref{hmu}) are odd under interchanging
$\a(i)$ with $\a(j)$; the last terms are even. Inserting this
expressions under the formula for $g^{\l,\mu}_{2n}$ only the odd terms
survive. Now the sum  in the eqs. for
$g^{\l,\mu}_{2n}$  is an equivalent way of
writing a pfaffian, thus
\begin{eqnarray*}
&&g_{2n}^{\l,\mu}=
\left[\prod\limits_{i<j}\sqrt{k}\,\sn i\a(ij)\right]^{-1}
\!\!\!\times{\rm Pfaffian}\left[
h_{\l,\mu}(i\a(i)-i\hat{\alpha}_{j})\right]
\end{eqnarray*}
Using the equivalence of the zeta function with the
logarithmic derivative of the $\th(u)$ function one sees that
the factors in square brackets are equivalent to the expressions on
page~\pageref{eq:two}, eqs.~(\ref{eq:hlm}). Using now the identity 
\begin{eqnarray*}
{\rm Pfaffian}\left(\sqrt{k}{\rm sn}(i\a(i)-i\a(j))\right)=
\prod\limits_{1\leq i< j\leq 2n}
\left(\sqrt{k}{\rm sn}(i\a(i)-i\a(j))\right).
\end{eqnarray*}
proved by use of Liouville's theorem, see e.g.~\cite{PT}, eq. (5.8).
This proves finally  eq.~(\ref{eq:two}).


\end{document}